\documentclass[a4paper,11pt]{article}
\pdfoutput=1

\usepackage{jinstpub}

\usepackage[load-configurations = abbreviations]{siunitx}
\usepackage{subfig}
\usepackage{supertabular,booktabs}
\usepackage[table,xcdraw]{xcolor}
\usepackage{adjustbox}

\newlength{\subcolumnwidth}
\newenvironment{subcolumns}[1][0.45\columnwidth]
 {\valign\bgroup\hsize=#1\setlength{\subcolumnwidth}{\hsize}\vfil##\vfil\cr}
 {\crcr\egroup}
\newcommand{\nextsubcolumn}[1][]{%
  \cr\noalign{\hfill}
  \if\relax\detokenize{#1}\relax\else\hsize=#1\setlength{\subcolumnwidth}{\hsize}\fi
}
\newcommand{\nextsubfigure}{\vfill}

\usepackage{lineno}

\title{\boldmath The influence of pixel cell layout on the timing performance of 3D sensors}

\author[a,1]{C. Lasaosa,\note{Corresponding author.}}
\author[a]{J. Duarte-Campderros,}
\author[a]{M. Fern\'andez,}
\author[a]{G. Gómez,}
\author[a]{I. Vila,}
\author[b]{S. Hidalgo,}
\author[b]{G. Pellegrini}
\affiliation[a]{Instituto de Física de Cantabria (CSIC-UC), Santander, Spain}
\affiliation[b]{Centro Nacional de Microelectrónica (IMB-CNM-CSIC), Barcelona, Spain}

\emailAdd{clara.lasaosa.garcia@cern.ch}

\abstract{Three-dimensional (3D) pixel sensors are a promising technology for implementing the 4D-tracking paradigm in high-radiation environments. Despite their advantages in radiation tolerance, 3D pixel sensors exhibit non-uniform electric and weighting fields that can degrade timing performance. This study explores the impact of pixel cell geometry on the timing characteristics of 3D columnar-electrode sensors fabricated by IMB-CNM, comparing square and hexagonal layouts. The sensors were characterized using the Two-Photon Absorption Transient Current Technique (TPA-TCT), providing high-resolution three-dimensional maps of the Time-of-Arrival (ToA) of charge carriers. Measurements at multiple depths and bias voltages reveal that the square geometry yields a more uniform temporal response compared to the hexagonal configuration. Additionally, a novel TPA-TCT-based method was introduced to determine the sensor jitter, relying on the analysis of the time difference between consecutive pulses in the TPA-TCT pulse train acquired under identical conditions. These findings underline the importance of pixel design optimization for future 4D-tracking detectors and confirm the TPA-TCT method as a powerful tool for detailed timing characterization.}


\proceeding{20$^{\text{th}}$ Trento Workshop on Advanced Silicon Radiation Detectors\\
  4 - 6 February 2025\\
  Trento, Italy}

\begin{document}
\maketitle
\flushbottom

\vspace{-0.8cm}
\section{Introduction}
\label{sec:intro}
\vspace{-0.2cm}

\begin{figure}[hbt!]
\vspace{-0.2cm}
\centering
\begin{subcolumns}[0.25\textwidth]
  \centering
  \subfloat[]{\hspace{0.9cm}\includegraphics[width=4.5cm]{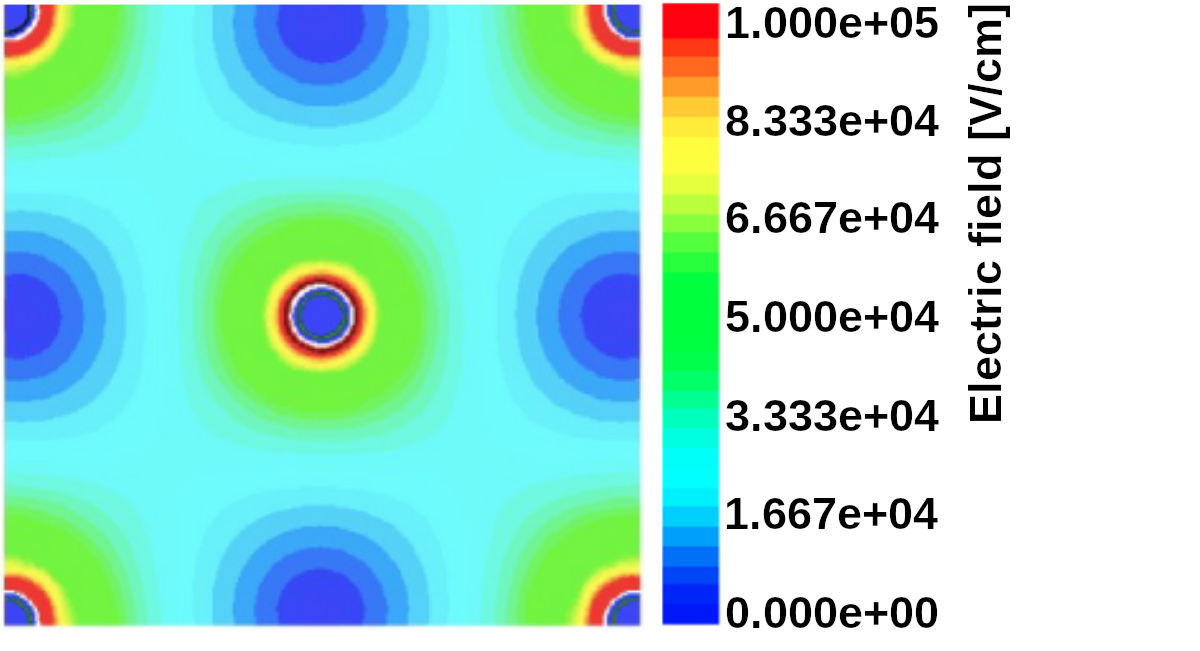}
  \label{fig:Efield_sq}}
  \nextsubfigure
  \centering
  \subfloat[]{\includegraphics[width=5.4cm]{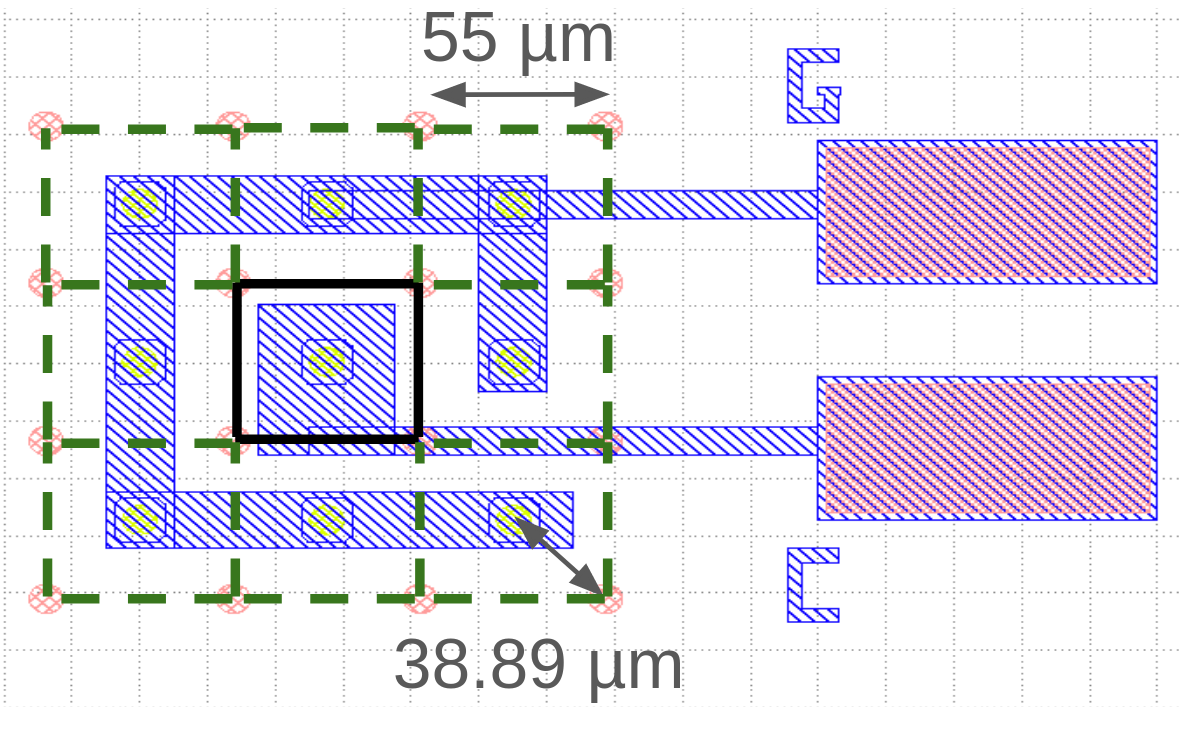}
  \label{fig:sample_sq}}
\nextsubcolumn[0.25\textwidth]
  \centering
  \subfloat[]{\hspace{1.25cm}\includegraphics[width=5cm]{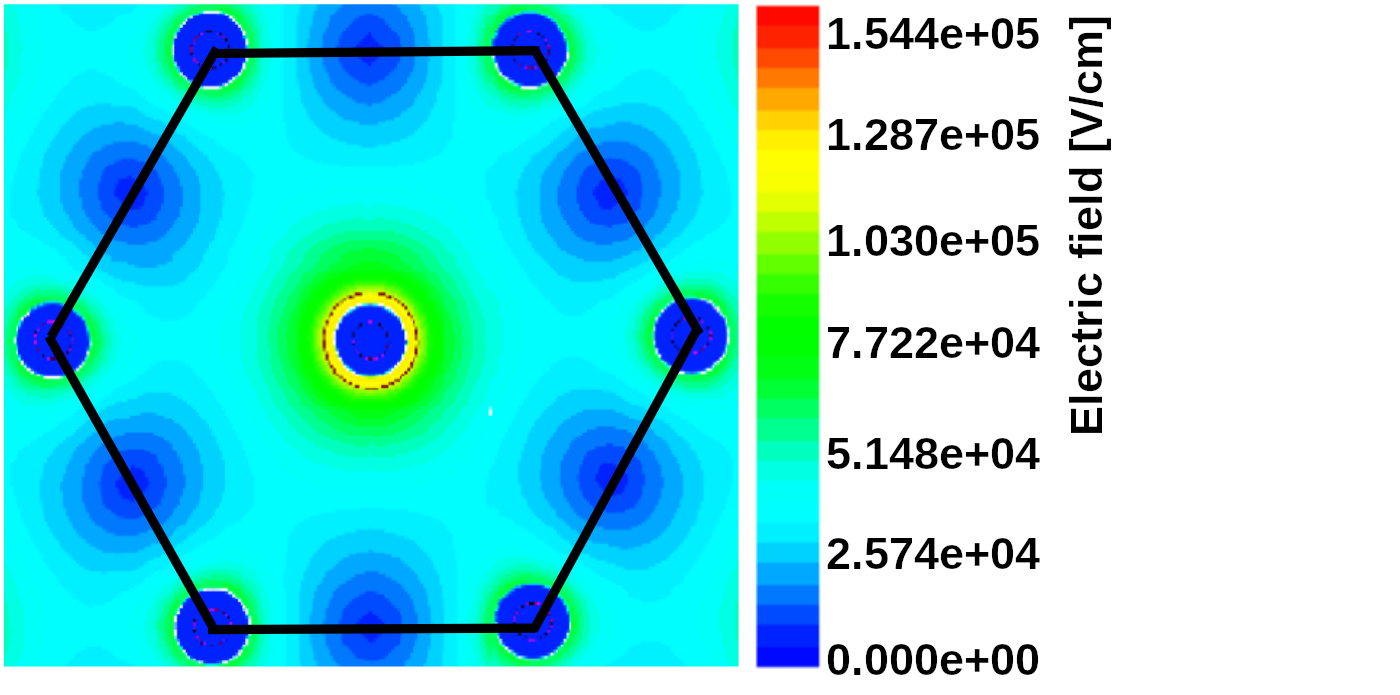}
  \label{fig:Efield_hx}}
  \nextsubfigure
  \centering
  \subfloat[]{\includegraphics[width=6.2cm]{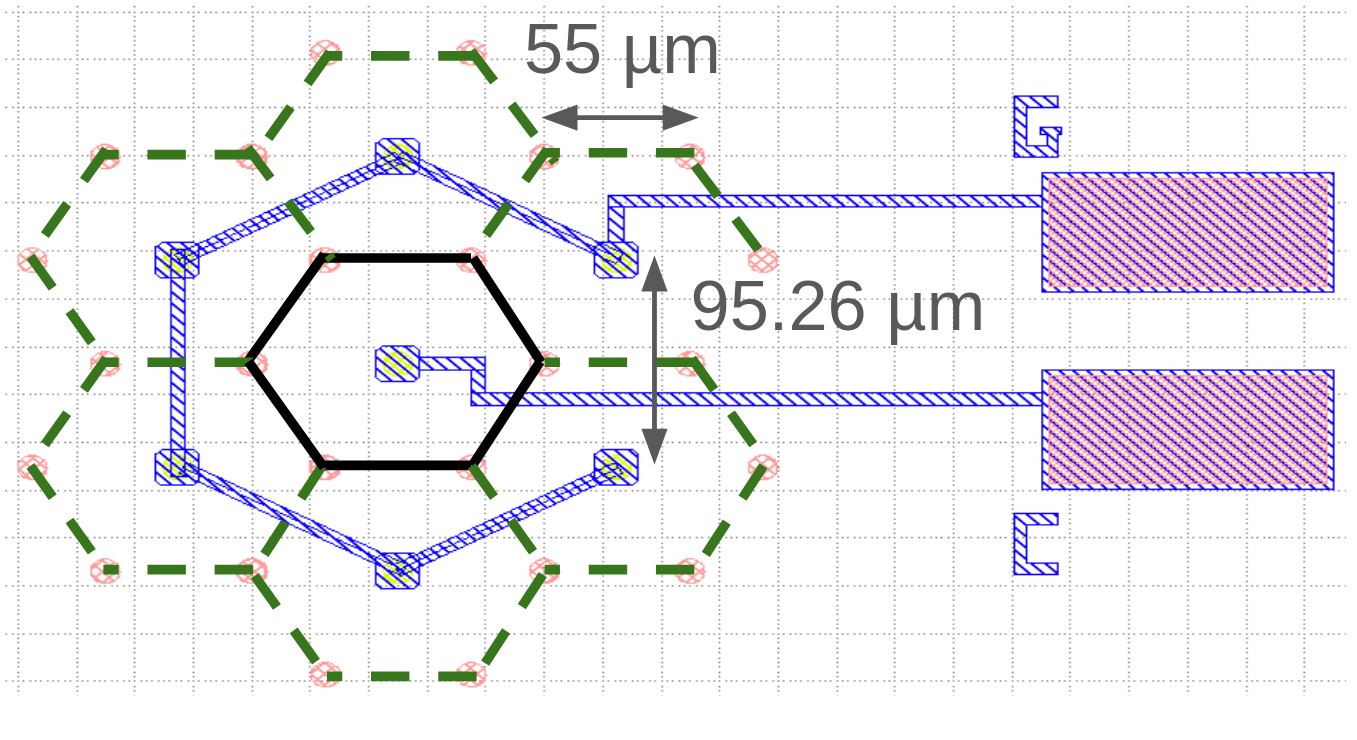}
  \label{fig:sample_hx}}
\nextsubcolumn[0.12\textwidth]
  \vspace{3.2cm}
  \subfloat[]{\hspace{0.1cm}\includegraphics[width=1.85cm]{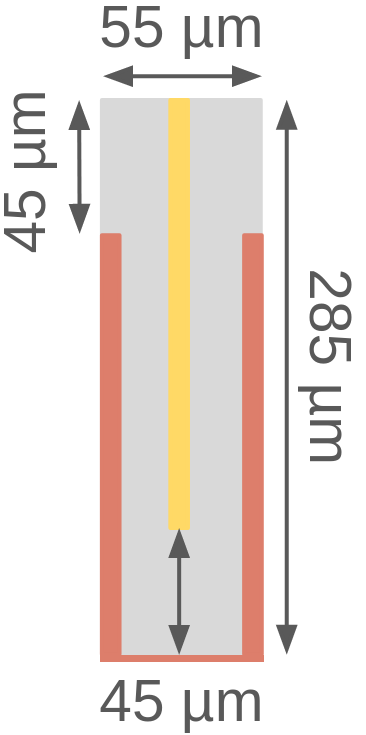}
  \label{fig:side_view}}
\end{subcolumns}
\vspace{-0.2cm}
\caption{Electric field computed using TCAD for (a) a square 3D pixel cell with a $\SI{55}{\micro m}$ pitch biased at $\SI{150}{V}$ and (c) a hexagonal 3D pixel cell with a $\SI{50}{\micro m}$ diameter biased at $\SI{100}{V}$. Images from~\cite{SquareSimulation, HexagonalSimulation}. Schematic top view of (b) square and (d) hexagonal 3D pixel sensors. The readout cell is outlined in black, with neighboring cells indicated in green. Metal pads are depicted in blue. (e) Schematic side view of the sensors, showing $p^{+}$ and $n^{+}$ electrodes in red and yellow, respectively.}
\label{fig:simulations-schematics}
\vspace{-0.25cm}
\end{figure}

Three-Dimensional (3D) pixel sensors~\cite{3Dsensors-Parker} are being explored for timing applications in extreme radiation environments, such as those anticipated in the proposed experiments at the Future Circular Collider or in potential upgrades of vertex detectors for the high-luminosity phase of the Large Hadron Collider~\cite{KRAMBERGER201926, DIEHL2024169517, LAI2020164491, ADDISON2023168392}. Detectors in these next-generation experiments require time resolutions on the order of tens or even units of picoseconds to accurately associate detected particles with their production vertex. However, considering timing applications, a major limitation of 3D pixel sensors compared to planar and avalanche sensor technologies is the non-uniformity of their electric and weighting fields, as shown in Figure~\ref{fig:simulations-schematics}. This leads to a variation in the signal detection time relative to the carrier generation time---commonly referred to as the \emph{Time-of-Arrival} (ToA)---depending on the particle impact position within the pixel cell. Accurately measuring the spread in ToA across the cell is therefore essential, as it directly affects the overall time resolution of the sensor.

The samples under test are unirradiated 3D columnar-electrode sensors with square and hexagonal pixel cells. These double-sided $n$-in-$p$ sensors, manufactured by IMB-CNM, have a thickness of \SI{285}{\micro m} and a pixel pitch of \SI{55}{\micro m}. Both $p^{+}$ and $n^{+}$ electrodes, each with a \SI{10}{\micro m} diameter, extend through the sensor, leaving a \SI{45}{\micro m} gap between their tip and the frontside or backside, respectively. Figure~\ref{fig:simulations-schematics} also includes the schematic layouts of the samples, where the central pixel cell is read out and the neighboring cells are shorted to ground.

\section{Experimental setup and methodology}
\label{sec:experimental-setup}
\vspace{-0.2cm}

The timing performance of these devices was evaluated with the Two-Photon Absorption Transient Current Technique (TPA-TCT) using the SSD laser facility at CERN~\cite{SebastianThesis}. The setup includes a fiber laser module (FYLA LFC1500X) to generate femtosecond pulses with a central wavelength of \SI{1550}{nm}, focusing optics to guide the light onto the sensor, and a readout system with a current-sensitive amplifier (CIVIDEC C2-TCT, \SI{2}{GHz}, \SI{46}{dB}) and a fast oscilloscope (Agilent DSO9254A, \SI{2.5}{GHz}, \SI{20}{GSa/s}). A neutral density filter is used to attenuate the pulse energy, and an acousto-optic modulator adjusts the repetition rate. In TPA-TCT, charge carriers are generated within a micrometric voxel centered at the beam focus, typically confined to an ellipsoidal excitation volume. This volume has a circular cross-section with a radius below \SI{1}{\micro m} and a longitudinal extent of less than \SI{10}{\micro m}. This highly localized excitation enables true three-dimensional mapping of the sensor's timing performance throughout its volume.

For the hexagonal sample, measurements were performed with beam illumination from the frontside. However, nearly the entire surface of the readout cell in the square sample is covered by metal pads (Figure~\ref{fig:sample_sq}), making frontside illumination prone to significant beam clipping. Locating the focus at a depth of \SI{285}{\micro m} in silicon requires, due to beam divergence, a metal-free region on the surface with a radius of approximately \SI{30}{\micro m}. To avoid interference from the metal pads, the square sample was illuminated from the backside, following the removal of the backside aluminum layer. Accordingly, the coordinate system used throughout this work defines the origin at the illuminated sensor surface---frontside for the hexagonal sample and backside for the square one.

Since beam clipping can also occur due to the presence of columnar electrodes, all measurements sensitive to beam intensity should be interpreted with caution. To mitigate systematic effects related to signal amplitude variations introduced by clipping, the ToA is extracted at a fixed fraction (50\%) of the signal amplitude, following the principle of the constant-fraction discrimination technique (CFD). It should be noted that the absolute value of the ToA is biased by an arbitrary time offset, primarily determined by the fixed delay between the laser trigger and the signal detection, which is dominated by the signal propagation time through the readout cables. 

For the ToA evaluation, the repetition rate was set to \SI{1}{kHz}, and each measurement averaged either \SI{128}{} or \SI{256}{} pulses to improve the signal-to-noise ratio (SNR) and suppress the jitter of both the sensor and the laser trigger. 

The TPA-TCT pulse train was employed for the first time to evaluate sensor time jitter with three-dimensional spatial resolution, extending the capabilities of the SPA-TCT method introduced in~\cite{LAI2020164491}. This advancement enables the local characterization of time jitter across the sensor volume, providing deeper insight into the spatial dependence of timing performance. These measurements were performed at the maximum repetition rate of \SI{8.2}{MHz} to record two pulses per waveform, both read out using the same scope channel to minimize jitter contributions from the readout system. The time spread, $\sigma$, is defined as the standard deviation obtained from a Gaussian fit to the distribution of the ToA differences between the two pulses, each extracted using the same CFD to correct for time walk. This approach allows the time jitter of the sample under test to be determined as $\sigma/\sqrt{2}$, without the need for an external reference sensor. The laser jitter between two consecutive pulses in the pulse train generated by the femtosecond fiber laser is negligible, owing to the high stability of the resonant cavity formed within the optical fiber. It should be noted that the quoted jitter values depend on the actual SNR during the measurement, and may therefore vary under different experimental conditions.

\section{Time-of-Arrival spread in 3D pixel sensors with square cell geometry}
\label{sec:results-square}
\vspace{-0.2cm}

\begin{figure}[hbt!]
\centering
\begin{subcolumns}[0.3\textwidth]
    \subfloat[]{\includegraphics[width=5.2cm]{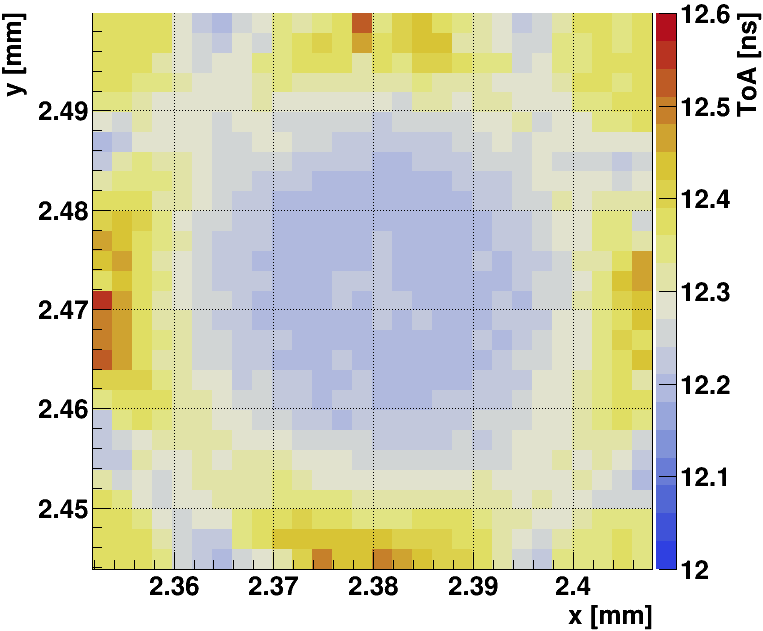}
    \label{fig:sq_map_gaplevel}}
\nextsubcolumn[0.3\textwidth]
    \subfloat[]{\includegraphics[width=5.2cm]{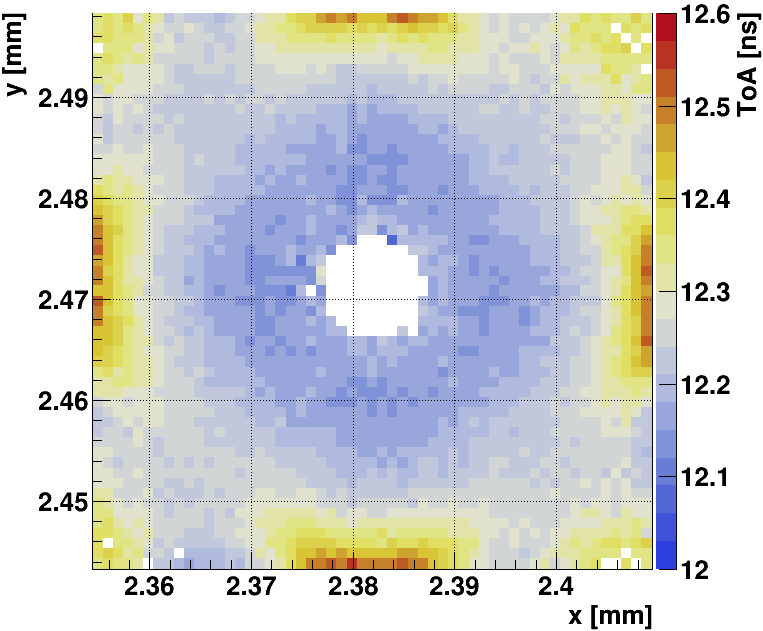}
    \label{fig:sq_map_deep}}
\nextsubcolumn[0.28\textwidth]
    \subfloat[]{\includegraphics[width=4.5cm, height=4.3cm]{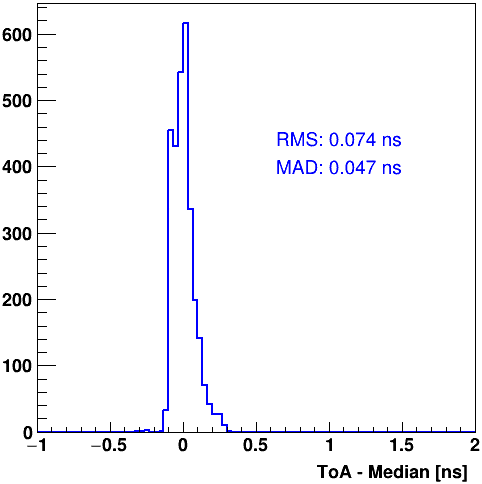}
    \label{fig:sq_1Ddist_deep}}
\end{subcolumns}
\vspace{-0.1cm}
\caption{ToA across the square pixel cell for a bias voltage of $\SI{20}{V}$, evaluated at two depths: (a) \mbox{$z_\mathrm{{Si}}\simeq\SI{23}{\micro m}$} and (b) \mbox{$z_\mathrm{{Si}}\simeq\SI{100}{\micro m}$}. (c) One-dimensional ToA distribution for \mbox{$z_\mathrm{{Si}}\simeq\SI{100}{\micro m}$}, corrected by the median. Key statistical parameters are indicated in blue.}
\label{fig:20V_sq}
\end{figure}

The ToA was evaluated across the square pixel cell at two different depths under a bias voltage of \SI{20}{V}, corresponding to an overdepleted sensor. Measurements were performed at approximately $z_{\text{Si}} \simeq \SI{100}{\micro\meter}$, where the $p^{+}$ and $n^{+}$ electrodes overlap, and at $z_{\text{Si}} \simeq \SI{23}{\micro\meter}$, a depth located beneath the $n^{+}$ column, roughly halfway between the column end and the sensor backside. Figures~\ref{fig:sq_map_gaplevel} and~\ref{fig:sq_map_deep} show the corresponding ToA maps at $z_{Si}\simeq\SI{23}{\micro m}$ and $z_{Si}\simeq\SI{100}{\micro m}$, respectively. As expected, signals within the inner radius of the $n^{+}$ column are only observed at $z_{\text{Si}} \simeq \SI{23}{\micro\meter}$.

Comparing the measured ToA maps with the simulated electric field map, the expected inverse relationship between ToA and electric field strength is observed. The highest ToA values are found at the pixel edges between the $p^{+}$ columns, where the electric field is low or even null, while lower values are observed toward the center of the pixel, where the electric field increases. Overall, a slightly lower ToA is observed in the central region at $z_{Si}\simeq\SI{100}{\micro m}$. Figure~\ref{fig:sq_1Ddist_deep} presents the one-dimensional ToA distribution at this depth, with a Median Absolute Deviation (MAD) of \SI{47}{ps} across the pixel cell. The MAD offers a more robust estimate of the spread compared to the RMS, as it is less sensitive to tails.

\begin{figure}[hbt!]
\centering
\begin{subcolumns}[0.3\textwidth]
    \subfloat[]{\includegraphics[width=4.5cm,height=4.21cm]{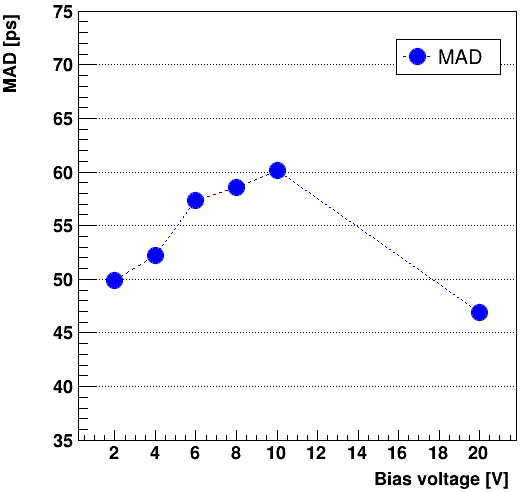}\label{fig:MAD_sq}}
    \nextsubfigure
    \subfloat[8 V]{\includegraphics[width=5cm]{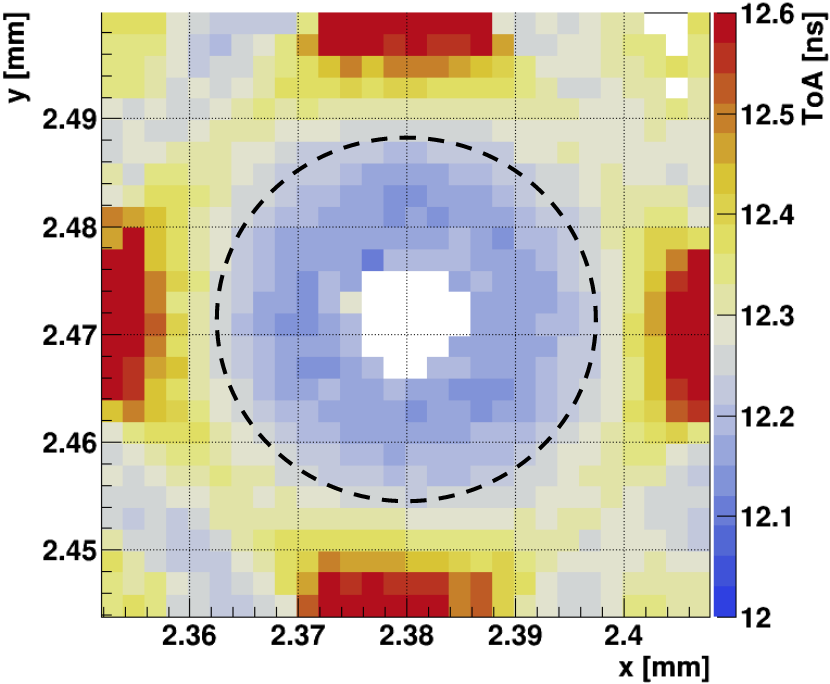}\label{fig:8V_sq}}
\nextsubcolumn[0.3\textwidth]
    \subfloat[20 V -- inner region]{\includegraphics[width=4.5cm,height=4.17cm]{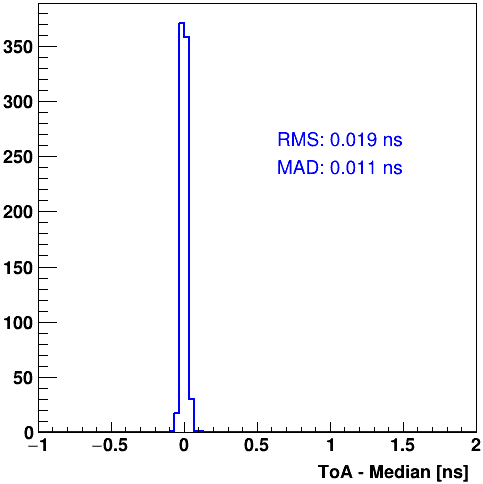}\label{fig:20V_sq-inner1D}}
    \nextsubfigure
    \subfloat[8 V -- inner region]{\includegraphics[width=4.5cm,height=4.17cm]{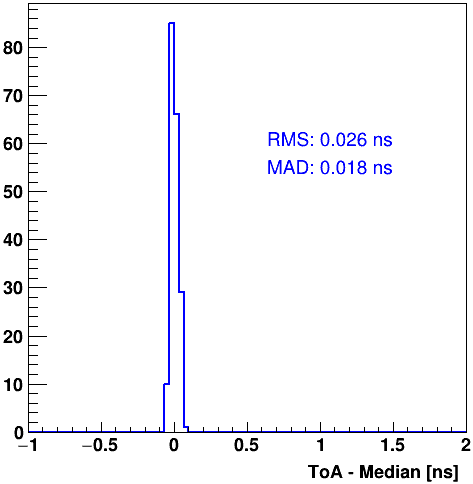}\label{fig:8V_sq-inner1D}}
\nextsubcolumn[0.3\textwidth]
    \subfloat[20 V -- outer region]{\includegraphics[width=4.5cm,height=4.17cm]{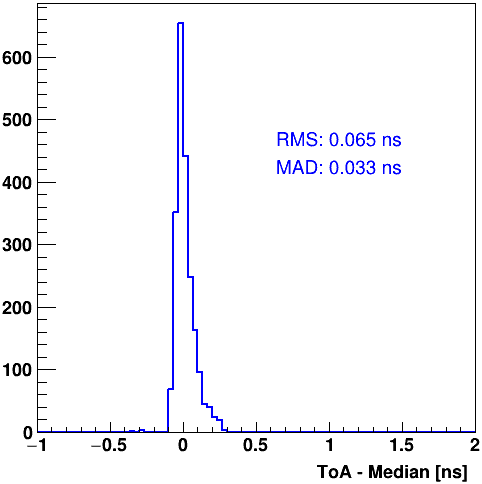}\label{fig:20V_sq-outer1D}}
    \nextsubfigure
    \subfloat[8 V -- outer region]{\includegraphics[width=4.5cm,height=4.17cm]{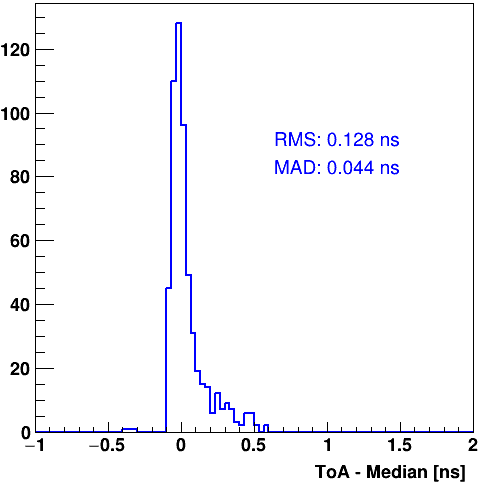}\label{fig:8V_sq-outer1D}}
\end{subcolumns}
\caption{(a) MAD of the one-dimensional ToA distribution as a function of bias voltage. (b) ToA across the square pixel cell at \mbox{$z_\mathrm{{Si}}\simeq\SI{100}{\micro m}$} for a bias voltage of \SI{8}{V}. Corresponding one-dimensional ToA distributions for the (d) inner and (f) outer regions of the pixel cell, defined by the black dashed line. Similar ToA distributions for a bias voltage of \SI{20}{V} are shown in (c) and (e).}
\label{fig:bias-voltage-scan_sq}
\end{figure}

ToA was also evaluated at several bias voltages to study its evolution with increasing electric field intensity. Measurements were conducted at $z_{Si}\simeq\SI{100}{\micro m}$ for bias voltages ranging from \SI{2}{V} to \SI{20}{V}. Figure~\ref{fig:MAD_sq} shows the MAD of the corresponding one-dimensional ToA distributions as a function of bias voltage. The spread in ToA is influenced by different transient behaviors in the inner and outer regions of the pixel cell, which respond differently to changes in bias voltage. Figures~\ref{fig:20V_sq-inner1D} and~\ref{fig:20V_sq-outer1D} present the one-dimensional ToA distributions for these regions at a bias voltage of \SI{20}{V} (ToA map in Figure~\ref{fig:sq_map_deep}), while Figures~\ref{fig:8V_sq-inner1D} and~\ref{fig:8V_sq-outer1D} show the same plots at \SI{8}{V} (ToA map in Figure~\ref{fig:8V_sq}). As the bias voltage increases, the inner and outer distributions tend to converge due to a more uniform electric field. This results in an artifact that produces an initial increase in MAD with bias voltage, followed by a decrease once the field becomes strong enough for the distributions of the inner and outer regions to significantly overlap.

\begin{figure}[hbt!]
\vspace{-0.2cm}
\centering
\begin{subcolumns}[0.3\textwidth]
    \subfloat[]{\includegraphics[width=5.4cm]{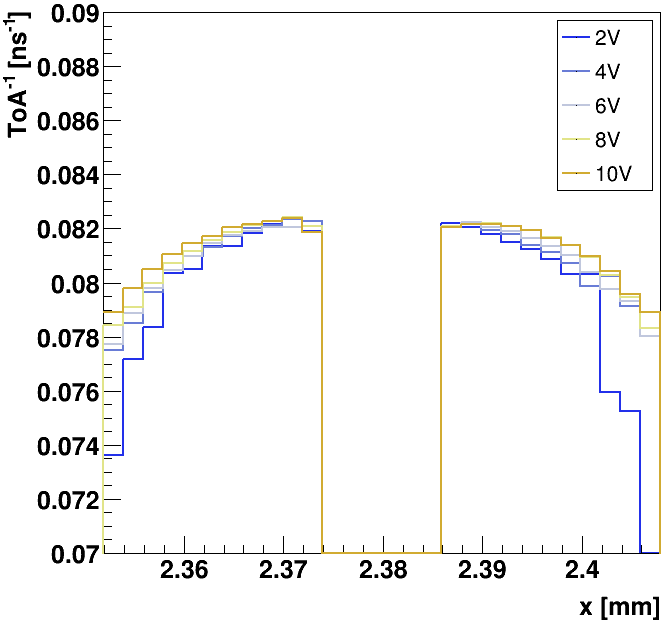}
    \label{fig:sq_profX}}
\nextsubcolumn[0.32\textwidth]
    \subfloat[]{\includegraphics[width=5.4cm]{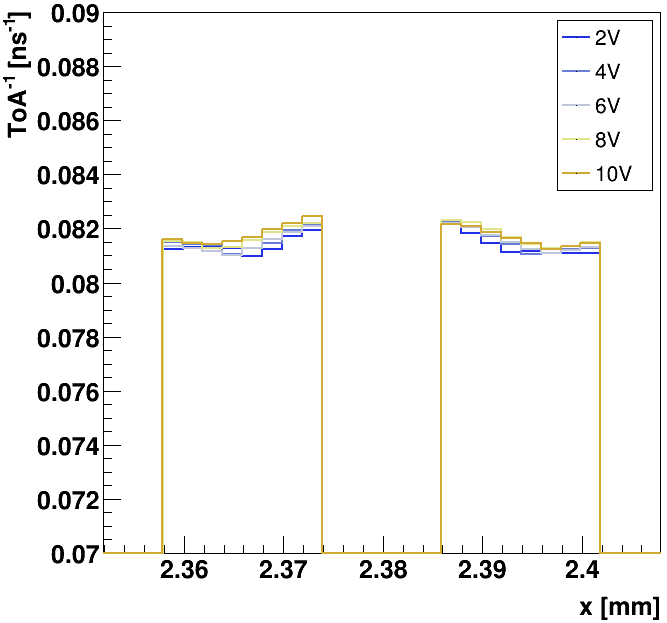}
    \label{fig:sq_profX_diagonal}}
\nextsubcolumn[0.15\textwidth]
    \subfloat[]{\includegraphics[width=2.2cm]{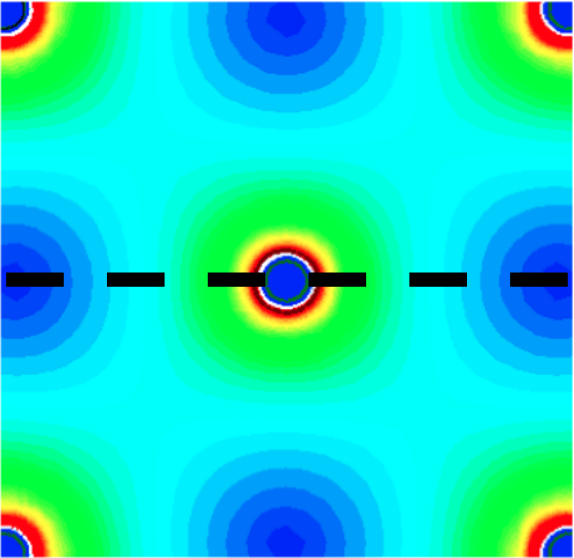}
    \label{fig:sq_profX_sim}}
    \nextsubfigure
    \vspace{-0.2cm}
    \subfloat[]{\includegraphics[width=2.2cm]{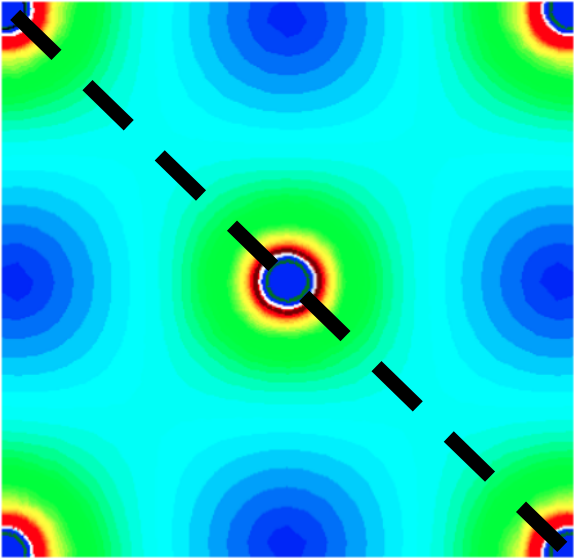}
    \label{fig:sq_profX_diagonal_sim}}
\end{subcolumns}
\vspace{-0.1cm}
\caption{Inverse ToA profiles for various bias voltages along (a) the horizontal axis and (b) the diagonal axis of the square pixel cell at \mbox{$z_\mathrm{{Si}}\simeq\SI{100}{\micro m}$}, as shown in (c) and (d), respectively.}
\label{fig:profiles_sq}
\end{figure}

Figures~\ref{fig:sq_profX} and~\ref{fig:sq_profX_diagonal} show the inverse ToA profiles extracted along the horizontal and diagonal axes, respectively, from the two-dimensional maps at various bias voltages. To reduce statistical fluctuations, the profiles were averaged over three adjacent bins centered on the axis of interest. The inverse of ToA provides a qualitative indication of the electric field strength. In the horizontal profile, the inverse ToA increases toward the $n^{+}$ column at the center of the cell, where the electric field is strongest. A similar trend is observed along the diagonal profile, although the variation is less pronounced due to the absence of the lowest field regions along this axis.

\begin{figure}[hbt!]
\centering
\begin{subcolumns}[0.24\textwidth]
    \subfloat[\hspace{0.5cm}Position 1]{\includegraphics[width=5.5cm]{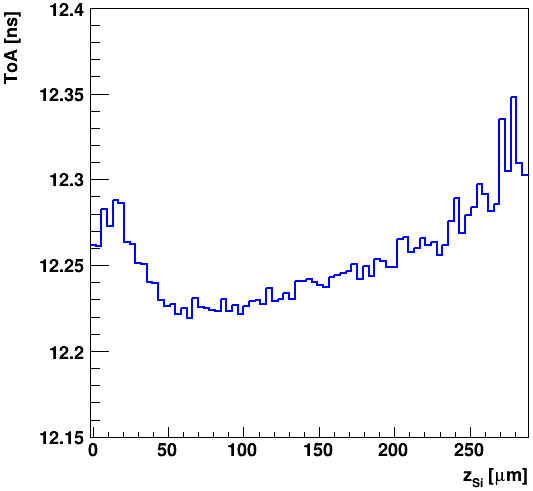}
    \label{fig:Zscan_pos1_sq}}
\nextsubcolumn[0.24\textwidth]
    \subfloat[\hspace{0.5cm}Position 2]{\includegraphics[width=5.4cm]{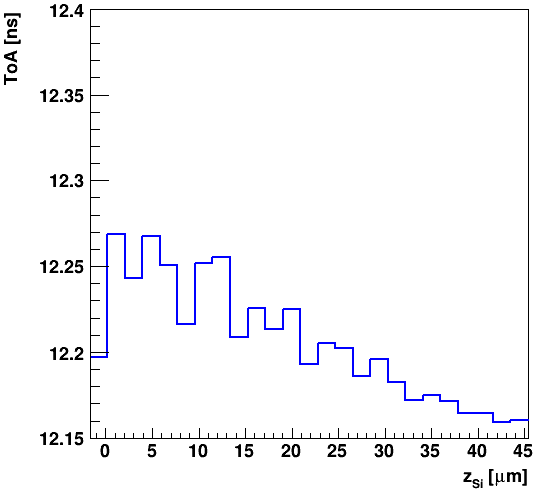}
    \label{fig:Zscan_pos2_sq}}
\nextsubcolumn[0.24\textwidth]
    \vspace{1.5cm}
    \subfloat[]{\includegraphics[width=4cm]{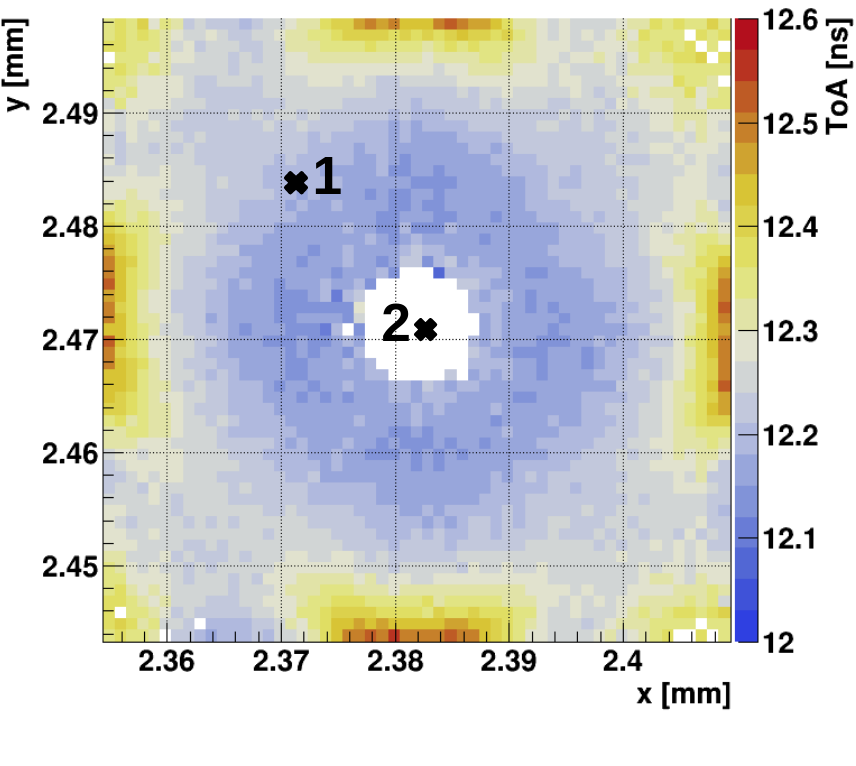}
    \label{fig:sq_positions}}
\end{subcolumns}
\vspace{-0.1cm}
\caption{(a) and (b) show the ToA as a function of depth at two different positions within the square pixel cell, as indicated in (c). The sensor was biased at $\SI{20}{V}$ during the measurements.}
\label{fig:Zscans_sq}
\end{figure}

Because the electrodes do not extend through the full sensor thickness, the electric field is not uniform along the depth of the sensor either. To study the dependence of ToA with depth, measurements were performed at the two positions indicated in Figure~\ref{fig:sq_positions} for a bias voltage of \SI{20}{V}. The first depth scan (Figure~\ref{fig:Zscan_pos1_sq}) was taken at a position approximately equidistant between the $n^{+}$ electrode and one of the $p^{+}$ electrodes. It shows an increase in ToA near both the frontside and backside, where there is no overlap between different types of electrodes. The second depth scan (Figure~\ref{fig:Zscan_pos2_sq}) was performed directly at the position of the $n^{+}$ column, providing significant signal amplitudes only along the gap. In this region, the ToA decreases as the focal point approaches the tip of the electrode, where the electric field reaches its maximum.

\section{Time-of-Arrival spread in 3D pixel sensors with hexagonal cell geometry}
\label{sec:results-hexagonal}
\vspace{-0.2cm}

\begin{figure}[hbt!]
\vspace{-0.3cm}
\centering
\begin{subcolumns}[0.3\textwidth]
    \subfloat[]{\includegraphics[width=5.2cm]{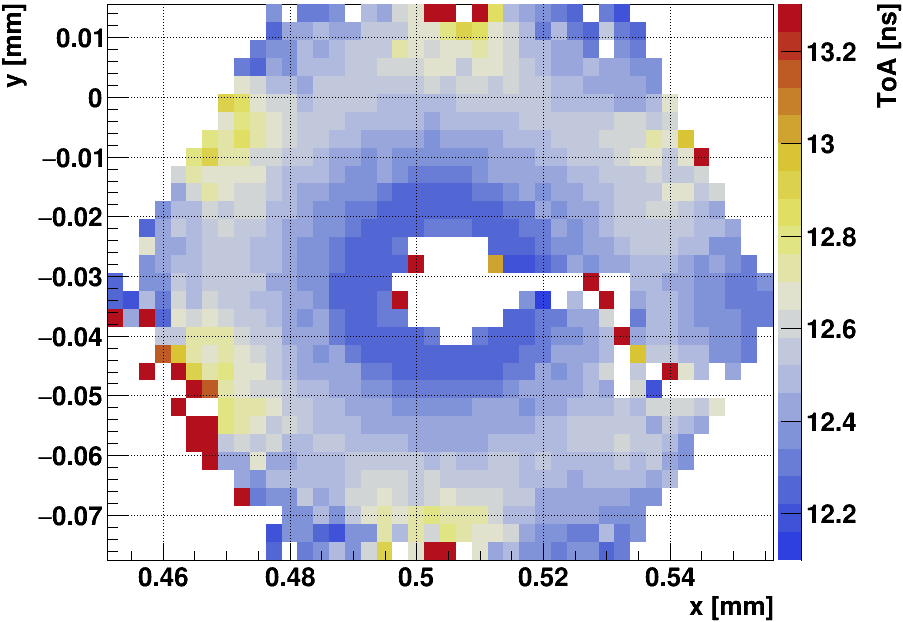}
    \label{fig:hx_map_surface}}
\nextsubcolumn[0.3\textwidth]
    \subfloat[]{\includegraphics[width=5.2cm]{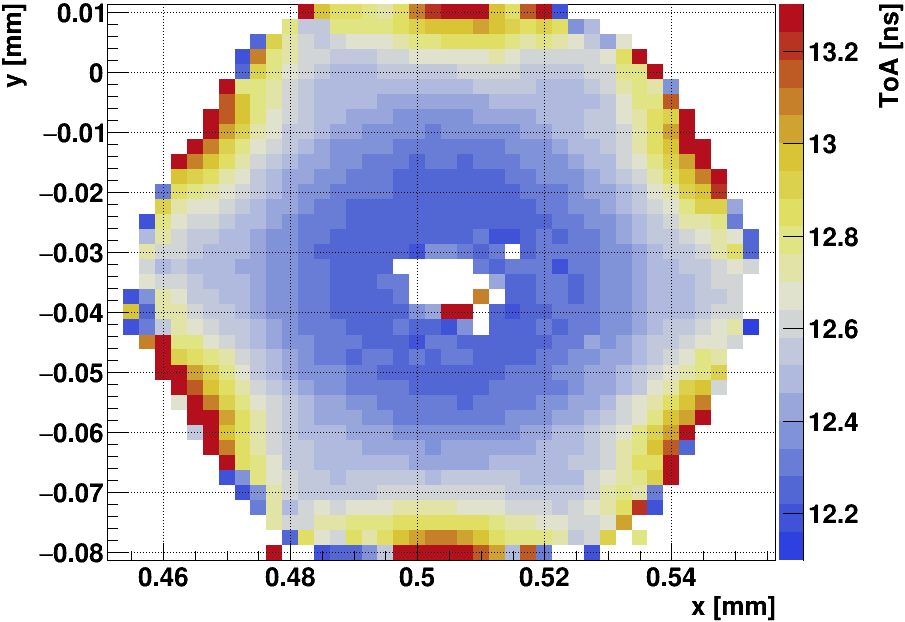}
    \label{fig:hx_map_deep}}
\nextsubcolumn[0.28\textwidth]
    \subfloat[]{\includegraphics[width=4.5cm]{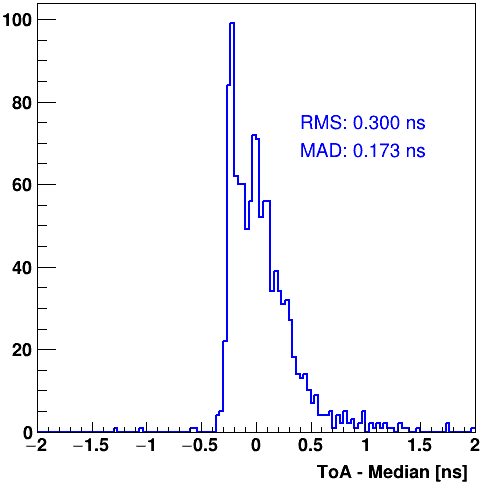}
    \label{fig:hx_1Ddist_deep}}
\end{subcolumns}
\vspace{-0.2cm}
\caption{ToA across the hexagonal pixel cell for a bias voltage of $\SI{20}{V}$, evaluated at two depths: (a) \mbox{$z_\mathrm{{Si}}\simeq\SI{0}{\micro m}$} and (b) \mbox{$z_\mathrm{{Si}}\simeq\SI{100}{\micro m}$}. (c) One-dimensional ToA distribution for \mbox{$z_\mathrm{{Si}}\simeq\SI{100}{\micro m}$}, corrected by the median. Key statistical parameters for both the original and corrected distributions are indicated in black and blue, respectively.}
\label{fig:20V_hx}
\vspace{-0.2cm}
\end{figure}

Similar studies were conducted with the hexagonal sample. The ToA across the pixel cell was evaluated at two depths for a bias voltage of \SI{20}{V}, as shown in Figure~\ref{fig:20V_hx}. Measurements were taken at approximately $z_{Si}\simeq\SI{100}{\micro m}$, where both $p^{+}$ and $n^{+}$ electrodes overlap, and at $z_{Si}\simeq\SI{0}{\micro m}$, where the presence of the readout metal track hides a small region of the map. As with the square sample, the lowest ToA appears around the central $n^{+}$ column, while the highest values are found near the pixel edges between $p^{+}$ columns. The corresponding one-dimensional ToA distribution at $z_{Si}\simeq\SI{100}{\micro m}$ is shown in Figure~\ref{fig:hx_1Ddist_deep}, revealing a significantly larger spread across the pixel cell than in the square sample, with a MAD of \SI{173}{ps}.

\begin{figure}[hbt!]
\centering
\begin{subcolumns}[0.3\textwidth]
    \subfloat[]{\includegraphics[width=5.4cm]{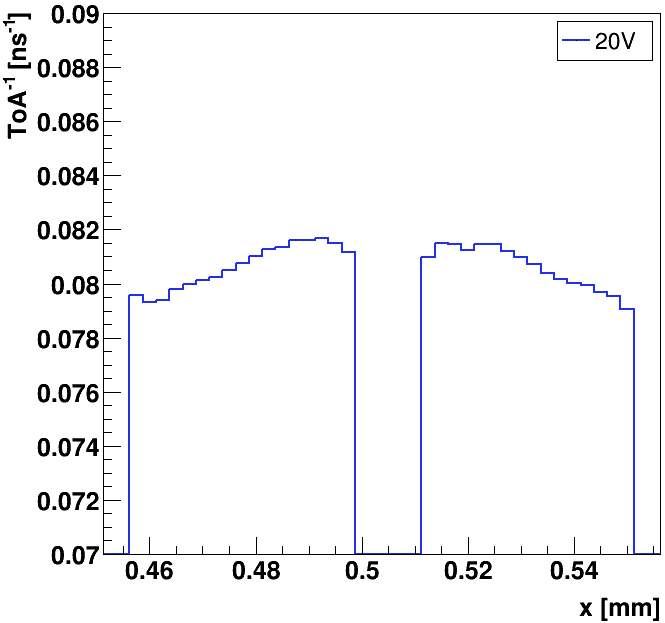}
    \label{fig:hx_prof_hor}}
\nextsubcolumn[0.32\textwidth]
    \subfloat[]{\includegraphics[width=5.48cm]{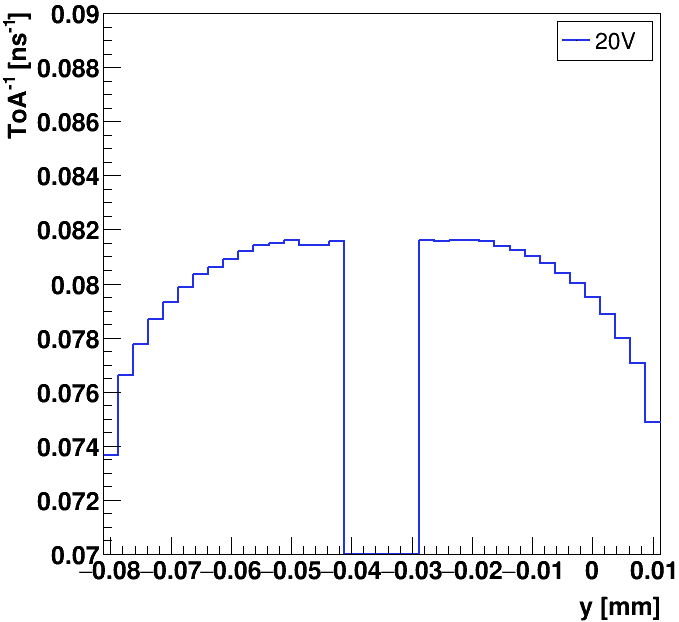}
    \label{fig:hx_prof_ver}}
\nextsubcolumn[0.15\textwidth]
    \subfloat[]{\includegraphics[width=2.2cm]{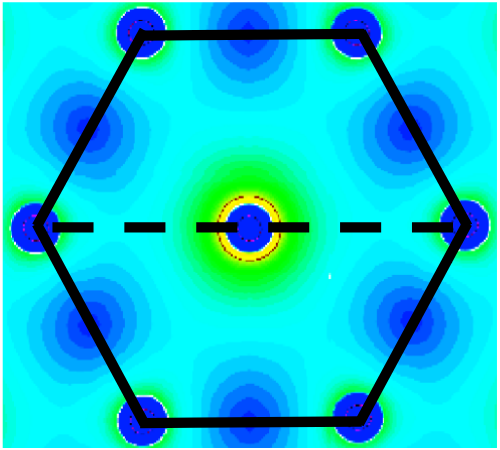}
    \label{fig:hx_prof_horizontal_sim}}
    \nextsubfigure
    \vspace{-0.2cm}
    \subfloat[]{\includegraphics[width=2.2cm]{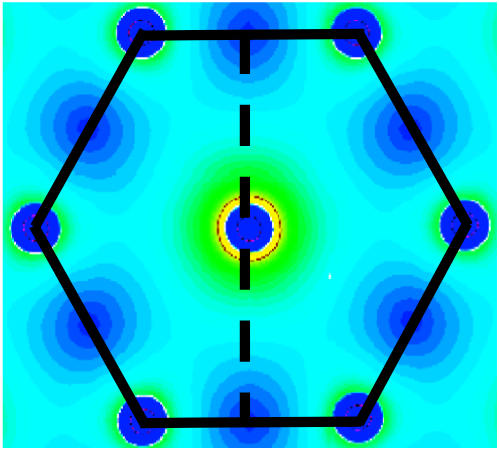}
    \label{fig:hx_prof_vertical_sim}}
\end{subcolumns}
\vspace{-0.1cm}
\caption{Inverse ToA profiles for a bias voltage of \SI{20}{V} along (a) the horizontal axis and (b) the vertical axis of the hexagonal pixel cell at \mbox{$z_\mathrm{{Si}}\simeq\SI{100}{\micro m}$}, as shown in (c) and (d), respectively.}
\label{fig:profiles-hx}
\end{figure}

The inverse ToA profiles along the horizontal and vertical axes of the map at $z_\mathrm{{Si}}\simeq\SI{100}{\micro m}$ for a bias voltage of \SI{20}{V} (Figure~\ref{fig:hx_map_deep}) are shown in Figures~\ref{fig:hx_prof_hor} and~\ref{fig:hx_prof_ver}, respectively. As with the square sample, the profiles were averaged over three adjacent bins centered on the axis of interest to improve statistics. Consistent with the expected electric field distribution, the inverse ToA increases toward the $n^{+}$ column at the center of the pixel. This variation is more pronounced along the vertical axis, which spans between two regions of low electric field.

\begin{figure}[hbt!]
\vspace{-0.1cm}
\centering
\begin{subcolumns}[0.3\textwidth]
    \subfloat[2 V]{\includegraphics[width=5cm]{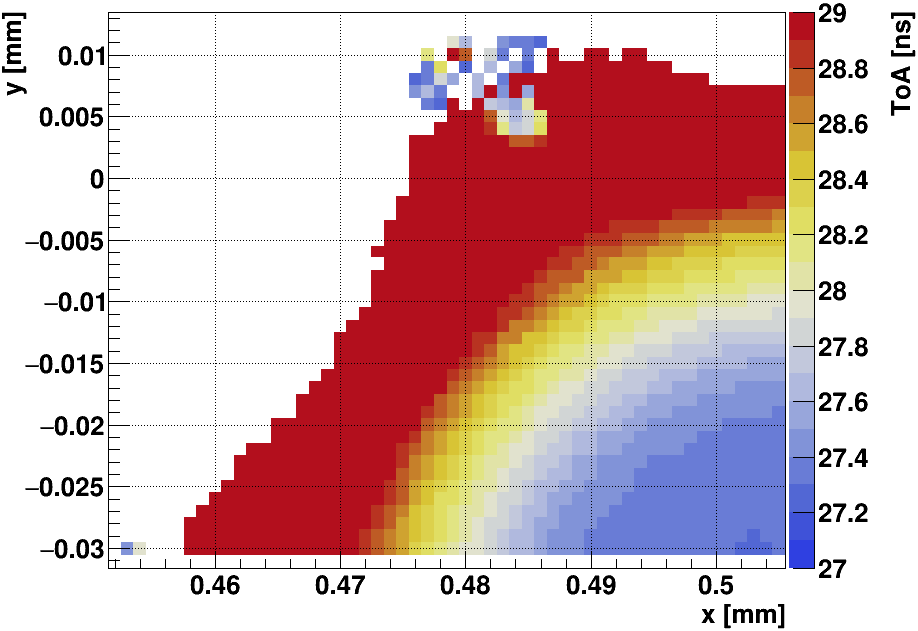}
    \label{fig:2V_hx}}
    \nextsubfigure
    \subfloat[10 V]{\includegraphics[width=5cm]{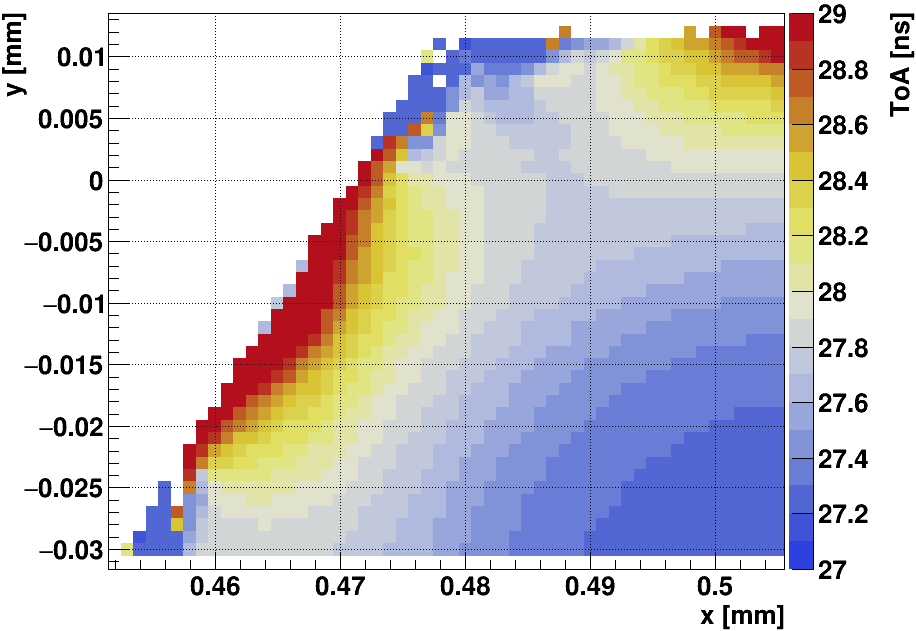}
    \label{fig:10V_hx}}
\nextsubcolumn[0.3\textwidth]
    \subfloat[4 V]{\includegraphics[width=5cm]{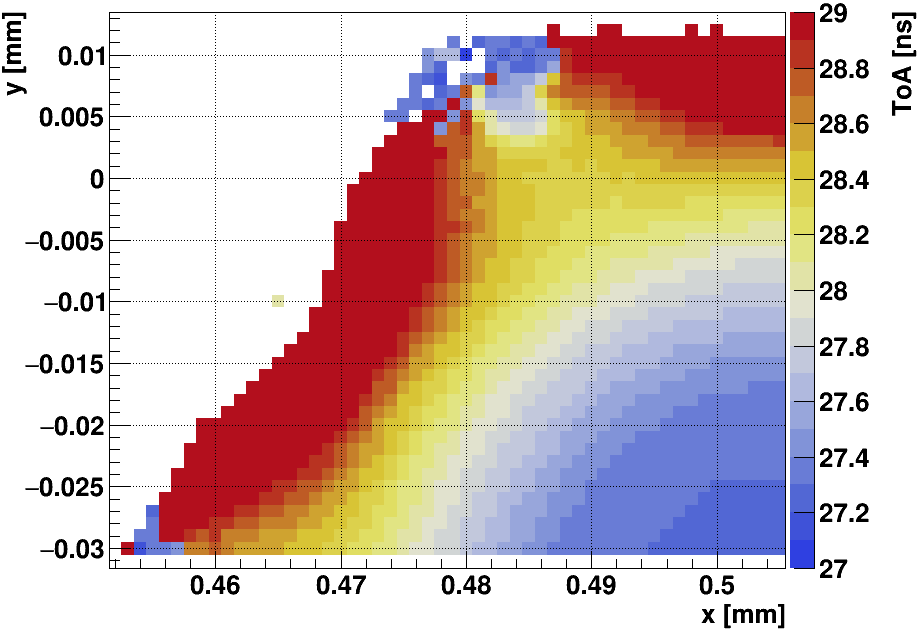}
    \label{fig:4V_hx}}
    \nextsubfigure
    \subfloat[14 V]{\includegraphics[width=5cm]{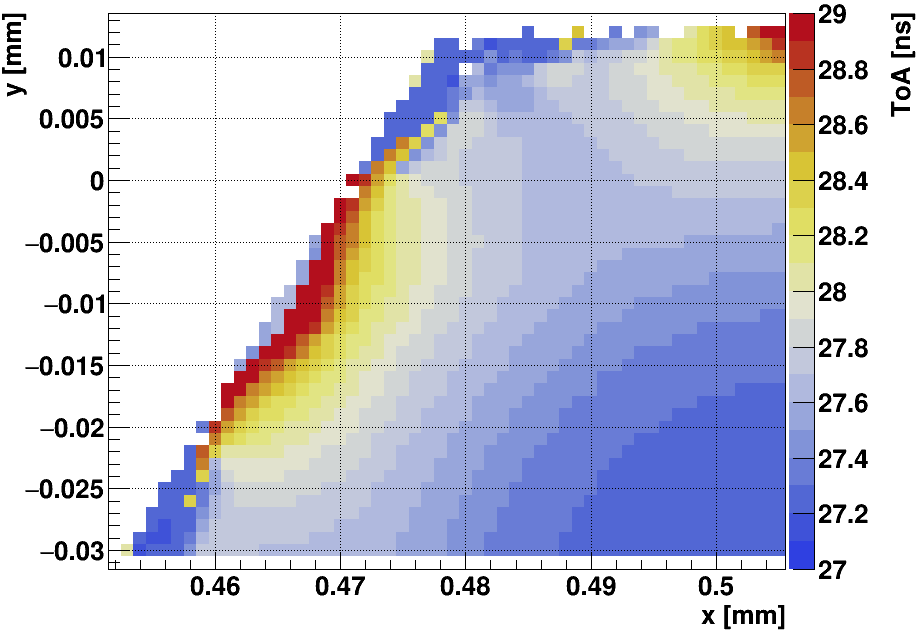}
    \label{fig:14V_hx}}
\nextsubcolumn[0.3\textwidth]
    \subfloat[6 V]{\includegraphics[width=5cm]{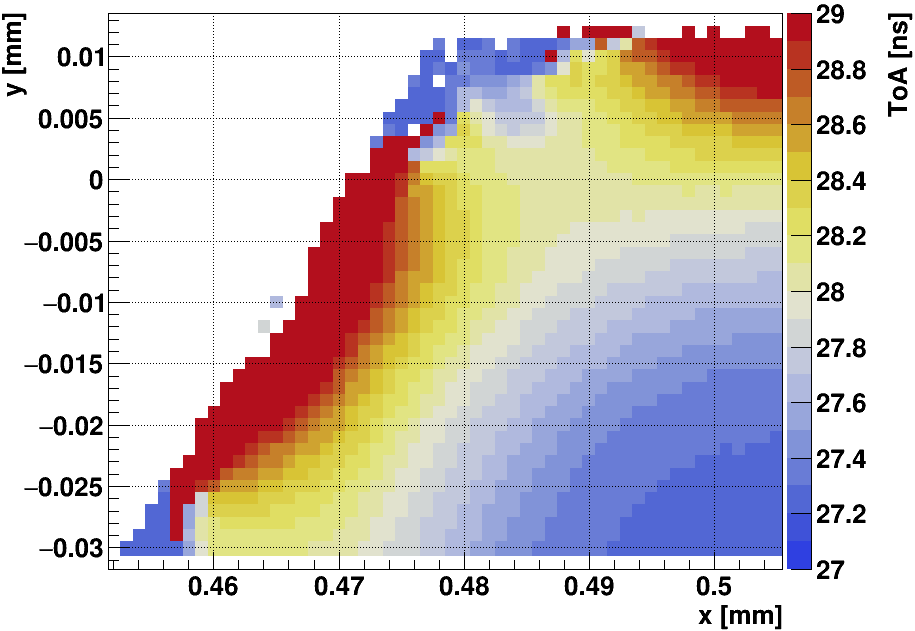}
    \label{fig:6V_hx}}
    \nextsubfigure
    \subfloat[]{\includegraphics[width=4.1cm,height=3.5cm]{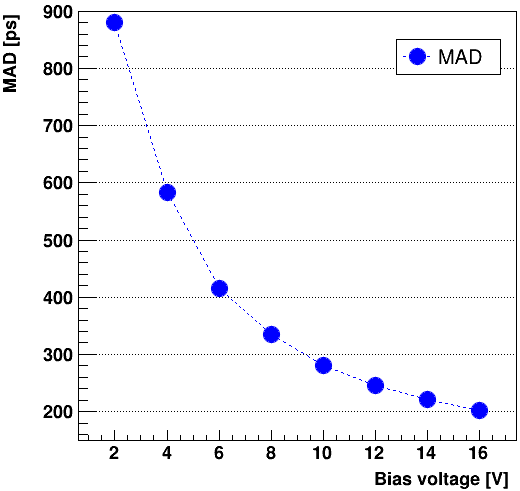}
    \label{fig:MAD_hx}}
\end{subcolumns}
\vspace{-0.1cm}
\caption{(a)-(e) ToA across one quarter of the hexagonal pixel cell at \mbox{$z_\mathrm{{Si}}\simeq\SI{100}{\micro m}$} for several bias voltages. (f) MAD of the associated one-dimensional ToA distributions as a function of bias voltage. The shift in the ToA magnitude is due to a change in the oscilloscope's time reference.}
\label{fig:hexagonal_quarter}
\end{figure}

Figure~\ref{fig:hexagonal_quarter} shows the ToA across one quarter of the hexagonal pixel cell at $z_{Si}\simeq\SI{100}{\micro m}$ for several bias voltages ranging from \SI{2}{V} to \SI{14}{V}. These maps illustrate the progressive depletion of the pixel cell and the changes in electric field intensity with increasing bias voltage. As shown in Figure~\ref{fig:MAD_hx}, the spread of the one-dimensional ToA distributions decreases with higher bias voltage, reflecting improved electric field uniformity.

\begin{figure}[hbt!]
\centering
\begin{subcolumns}[0.24\textwidth]
    \subfloat[\hspace{0.5cm}Position 1]{\includegraphics[width=5.5cm]{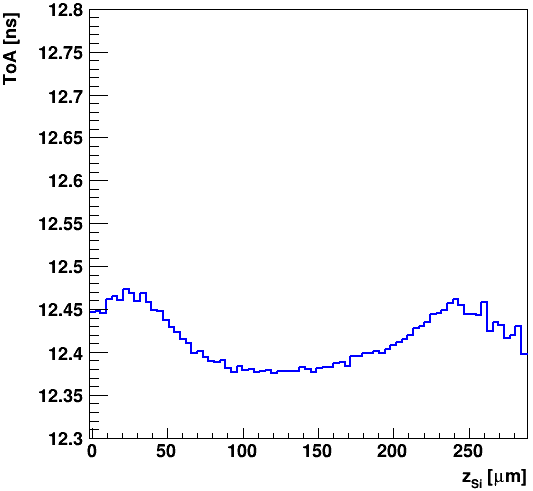}
    \label{fig:Zscan_pos1}}
\nextsubcolumn[0.24\textwidth]
    \subfloat[\hspace{0.5cm}Position 2]{\includegraphics[width=5.4cm]{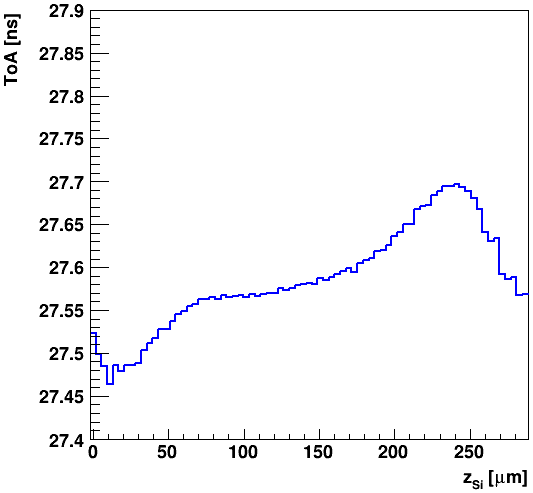}
    \label{fig:Zscan_pos2}}
\nextsubcolumn[0.24\textwidth]
    \vspace{2.1cm}
    \subfloat[]{\includegraphics[width=4cm]{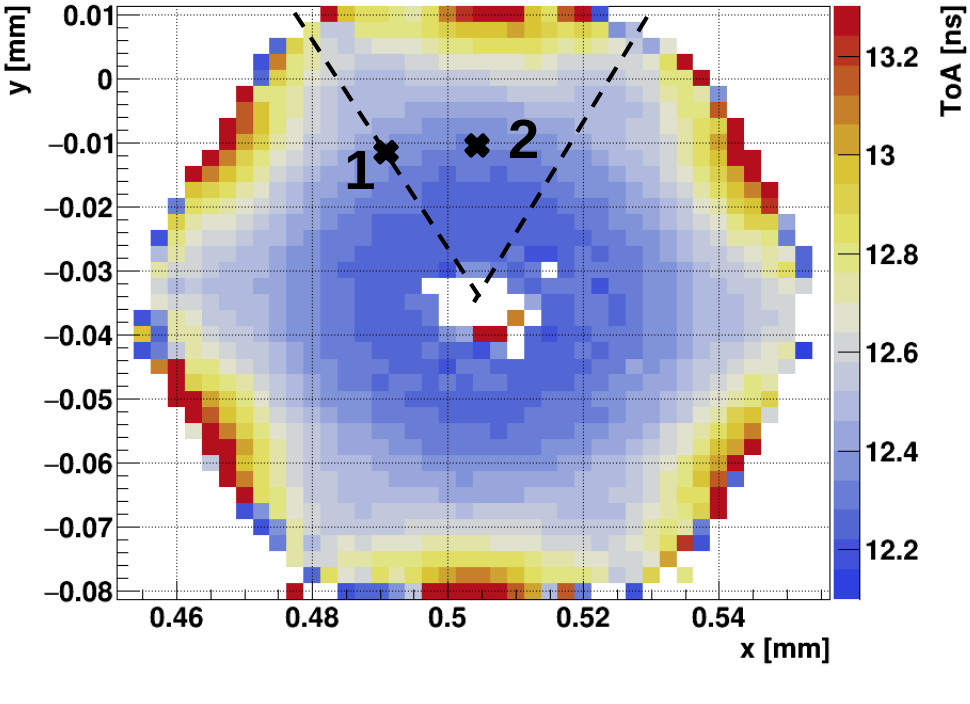}
    \label{fig:hx_positions}}
\end{subcolumns}
\vspace{-0.1cm}
\caption{(a) and (b) show the ToA as a function of depth at two positions within the hexagonal pixel cell, as indicated in (c). The sensor was biased at \SI{20}{V} for these measurements. The shift in the ToA magnitude is due to a change in the oscilloscope's time reference.}
\label{fig:Zscans_hx}
\end{figure}

The depth dependence of the ToA was also investigated for the hexagonal sample, as shown in Figure~\ref{fig:Zscans_hx}. Measurements were performed for a bias voltage of \SI{20}{V} at the two positions indicated in Figure~\ref{fig:hx_positions}. The resulting profiles differ between the two positions, particularly near the frontside and backside, where the $p^{+}$ and $n^{+}$ electrodes do not overlap.

\section{Sensor time jitter determination in 3D pixels with hexagonal cell geometry}\label{sec:jitter-hexagonal}
\vspace{-0.2cm}

For the hexagonal sample, the sensor time jitter was evaluated at the first position indicated in Figure~\ref{fig:hx_positions}, following the procedure described in Section~\ref{sec:experimental-setup}. The measurement was performed at a bias voltage of \SI{20}{V}, with laser intensity equivalent to approximately \SI{9.5}{} minimum ionizing particles. The sensor time jitter was assessed for several CFD thresholds, achieving the best jitter at the specified position---\SI{22}{ps}---with a 65\% CFD. 

\section{Conclusions}\label{sec:conclusions}
\vspace{-0.2cm}

Pixel cell geometry was found to significantly influence timing performance. The square cell layout exhibits a more homogeneous ToA response than the hexagonal layout, with a smaller spread of approximately~\SI{45}{ps} compared to~\SI{175}{ps} at a bias voltage of \SI{20}{V}, in both cases under fully depleted conditions. It is worth noting that the distance between electrodes of opposite polarity is larger in the hexagonal cell. As a result, the reduced inter-electrode electric field in the hexagonal geometry contributes for the larger observed spread in ToA. The inverse ToA maps qualitatively match electric field profiles from TCAD simulations, providing a practical proxy for visualizing non-uniformities in the internal field structure.

Moreover, a novel method was developed to estimate the sensor time jitter using the TPA-TCT pulse train. By analyzing time differences between consecutive laser pulses under identical conditions, this technique yielded sensor time jitter values down to~\SI{22}{ps}, offering a reliable and reproducible way to characterize time performance.

The use of TPA-TCT with femtosecond infrared laser pulses enabled high-resolution, three-dimensional mapping of the timing response in 3D pixel sensors. This level of spatial and temporal resolution cannot be achieved using radioactive sources or standard single-photon TPA techniques, highlighting the unique diagnostic power of femtosecond TPA-TCT for precision studies of charge collection dynamics in complex sensor geometries.

\acknowledgments
\vspace{-0.2cm}

The authors gratefully acknowledge the support and access to the Solid State Detector (SSD) laboratory at CERN, which enabled the TPA-TCT measurements, as well as the support of the RD50 Collaboration. This work was partially supported by the AIDAInnova project, funded by the European Union’s Horizon 2020 Research and Innovation Programme under Grant Agreement No.~101004761. Additional funding was provided through Grants PID2023-148418NB-C41 and PID2023-148418NB-C42, financed by MCIN/AEI/10.13039/501100011033, and by the Complementary Plan in Astrophysics and High-Energy Physics (CA25944), project C17.I02.P02.S01.S03 CSIC, supported by the Next Generation EU funds, RRF and PRTR mechanisms, and the Government of the Autonomous Community of Cantabria. This work has also received partial support from the Spanish Ministry of Science and Innovation through the Ramón y Cajal Programme [RYC2023-044327-I].

\section*{Author Contributions}
\vspace{-0.2cm}

The contributions of each author are described according to the CRediT (Contributor Roles Taxonomy) classification scheme. \textbf{C. Lasaosa:}  Software, Validation, Formal analysis, Investigation, Data curation, Writing – original draft. \textbf{J. Duarte-Campderros:}  Writing – review, Supervision. \textbf{M. Fernández:} Software, Validation, Investigation, Data curation, Writing – review. \textbf{G. Gómez:}  Writing – review, Supervision, Funding acquisition. \textbf{I. Vila:} Conceptualization, Writing – review \& editing, Supervision, Project administration, Funding acquisition. \textbf{S. Hidalgo}: Resources. \textbf{G. Pellegrini}: Resources.

\bibliography{TREDI2025_CLG}

\end{document}